\documentclass{article}  
\usepackage{jamaica04}
\frompage{000} \topage{000}                                              
\def\rightmark{Proton - Lambda correlations in Au-Au Collisions}
\def\leftmark{G. Renault et al.}
 
\title{Proton - Lambda correlations in Au-Au Collisions\\ 
at  $\sqrt{s_{NN}} = 200$ GeV from the STAR experiment\\} 
\authors{
{G. Renault$^1$ for the STAR Collaboration$^{2}$ %
}\\[2.812mm]
{\normalsize
\hspace*{-8pt}$^1$ SUBATECH,\\
Laboratoire de Physique Subatomique et des Technologies Associ\'ees\\ 
University of Nantes - IN2P3/CNRS - Ecole des Mines de Nantes \\
4 rue Alfred Kastler, F-44307 Nantes Cedex 03, France\\[0.2ex]
\hspace*{-8pt}$^2$ http://www.star.bnl.gov/central/collaboration/authorList.php\\
}} 
\abstract{The space-time evolution of the
source of particles formed in the collision of nuclei
can be studied through particle correlations. 
The STAR experiment is dedicated to study ultra-relativistic
heavy ions collisions and allows to measure non-identical
strange particle correlations. The source size can be 
extracted by studying $p-\Lambda$, $\overline{p}-\overline{\Lambda}$,
$\overline{p}-\Lambda$ and $p-\overline{\Lambda}$ correlation functions.
Strong interaction potential has been studied for these systems
using an analytical model.
Final State Interaction (FSI) parameters have been determined and has shown
a significant annihilation process present in $\overline{p}-\Lambda$ and
$p-\overline{\Lambda}$ systems not present in $p-\Lambda$ and
$\overline{p}-\overline{\Lambda}$.
}
\begin{document} 
\maketitle
\setcounter{page}{1}
\section{Introduction}\label{intro}
 
Non-identical particles are correlated due to 
final state Coulomb and nuclear interactions \cite{1}.
Contrary to $p-\Lambda$ \cite{2,5,6},
the nuclear FSI (Final State Interaction) 
of $\overline{p}-\Lambda$, $p-\overline{\Lambda}$
and $\overline{p}-\overline{\Lambda}$, is still unknown.
In this paper, data from the STAR experiment are shown and
the Lednick\'y \& Lyuboshitz model \cite{3}
is used to analyse experimental correlation functions \cite{10}.
The STAR detector (a Solenoid Tracker At RHIC), installed 
at RHIC (Relativistic Heavy Ion Collider), 
allows the reconstruction of the 
particles produced during the Au+Au collisions at 200 GeV
per nucleon pair in the center of mass.
 
\section{Experimental correlation functions}\label{techno}  
The particles are measured in Au-Au collisions at 
 $\sqrt{s_{NN}} = 200$ GeV using the Time Projection Chamber
(TPC). Central events accounting for 10\% of the total 
cross section are selected.

The relevant variable is the momentum of one 
of the particles in the pair rest frame called here 
$\overrightarrow{k}^*$.
The non-correlated background is constructed by mixing events 
with primary vertex separated from each other by less than 10 cm.
The correlation function has been extracted by 
constructing the ratio of two distributions.
The numerator is the $|\overrightarrow{k^{*}}|$ distribution
of pairs from the same event. 
The denominator 
is the $|\overrightarrow{k^{*}}|$ distribution of pairs from mixed events. 


Protons and anti-protons
are selected using their specific energy loss ($dE/dx$).
This selection limits the acceptance of particles to 
the transverse momentum range of 0.4-1.1 GeV/c in the 
rapidity interval $|Y|<0.5$.

Lambdas (anti-lambdas) are reconstructed through the decay channel
$\Lambda \rightarrow \pi + p$ 
($\overline{\Lambda} \rightarrow \pi + \overline{p}$), 
with a corresponding branching ratio of 64\%.
Pions and protons are selected using their 
specific energy loss. In addition some geometrical cuts
are applied, giving a lambda/anti-lambda purity sample of 86\%,
the remaining 14\% representing the combinatory background.
Only lambdas/anti-lambdas in the rapidity 
range $|Y|<1.5$ are selected.
Due to the acceptance of the detector, the transverse momentum 
range is 0.3-2.0 GeV/c.


The contamination and the feed-down have been studied
in order to estimate the purity (Eq. (\ref{Purity})) 
of $p,\overline{p},\Lambda$ and $\overline{\Lambda}$
as a function of the transverse momentum ($p_{t}$).
The purity is defined as the product of the probability of identification (Pid) 
multiplied by the fraction of primary particles (Fp).
\begin{eqnarray}
\label{Purity}
\textrm{Purity} (p_{t}) = \textrm{Pid} (p_{t}) * \textrm{Fp} (p_{t})
\end{eqnarray}
The probability of identification has been estimated as a function of $k^{*}$
for charged particles.
Identified protons (anti-protons) from the selected sample 
account for $76.5\pm2\%$ ($74\pm2\%$).
For lambdas (anti-lambdas) the probability of
identification correspond to the signal over noise value, 
it is independent of $|\overrightarrow{k^{*}}|$: $86.54\pm0.04\%$ ($86.13\pm0.04\%$).

The feed-down estimation has been done for 
$p$, $\overline{p}$, $\Lambda$ and $\overline{\Lambda}$
as a function of the transverse momentum ($p_{t}$) in order 
to take into account the $|\overrightarrow{k^{*}}|$ dependence of the purity.
Combined results from STAR \cite{11,12,13,14,15,16} and 
predictions from the
thermal model \cite{9} have been used. 
The approximations done by estimating the purity,
raise the problem of considerable uncertainties on extracted values for FSI parameters 
and radii.     

The calculated feed-down
leads to an estimated purity of 54\% for primary protons ($<p_{t}>=0.70$ GeV/c). 
Most of the secondary protons come
from lambda decay and represent 35\% of the protons
used to construct the correlation function. Other sources
of contamination of protons are provided by decay products 
of $\Sigma^{+}$ and 
pions interacting with matter, which represent respectively
10\% and 1\% of the sample.

The feed-down study for anti-protons ($<p_{t}>=0.73$ GeV/c),
leads to an estimated purity of 56\% for primary anti-protons. 
Most of the secondary anti-protons come
from anti-lambda decay and represent 32\% of the anti-protons
used to construct the correlation function.
For anti-protons, the additional source
of contamination, which is the decay product of $\overline{\Sigma^{+}}$,
represents 12\%.

The sample of lambdas (anti-lambdas) includes secondary particles 
such as decay products of
$\Xi, \Xi^{0}, \Sigma^{0}$ 
($\overline{\Xi}, \overline{\Xi^{0}}, \overline{\Sigma^{0}}$).
The fractions of lambdas ($<p_{t}>=1.20$ GeV/c) 
and anti-lambdas ($<p_{t}>=1.23$ GeV/c) 
coming from the primary vertex have been estimated 
at 44\%. 

The pair purity plays a crucial role in the correlation study.
The estimated value of the pair purity for $p-\Lambda$,
 $\overline{p}-\overline{\Lambda}$, $p-\overline{\Lambda}$ and 
$\overline{p}-\Lambda$ systems is 15\%.
The contamination tends to reduce the correlation strength.
Raw data has been corrected for purity using the following method:
\begin{eqnarray} 
C_{true}(k^*) = \frac{C_{measured(k^*)}-1}{\textrm{PairPurity}(k^*)} + 1
\end{eqnarray}
where PairPurity represents the product of the two value of purity of particles,
$C_{true}(k^*)$ represents the corrected correlation function and
$C_{measured(k^*)}$ represents the measured one.

\begin{figure}[htb!]
\center
\insertplot{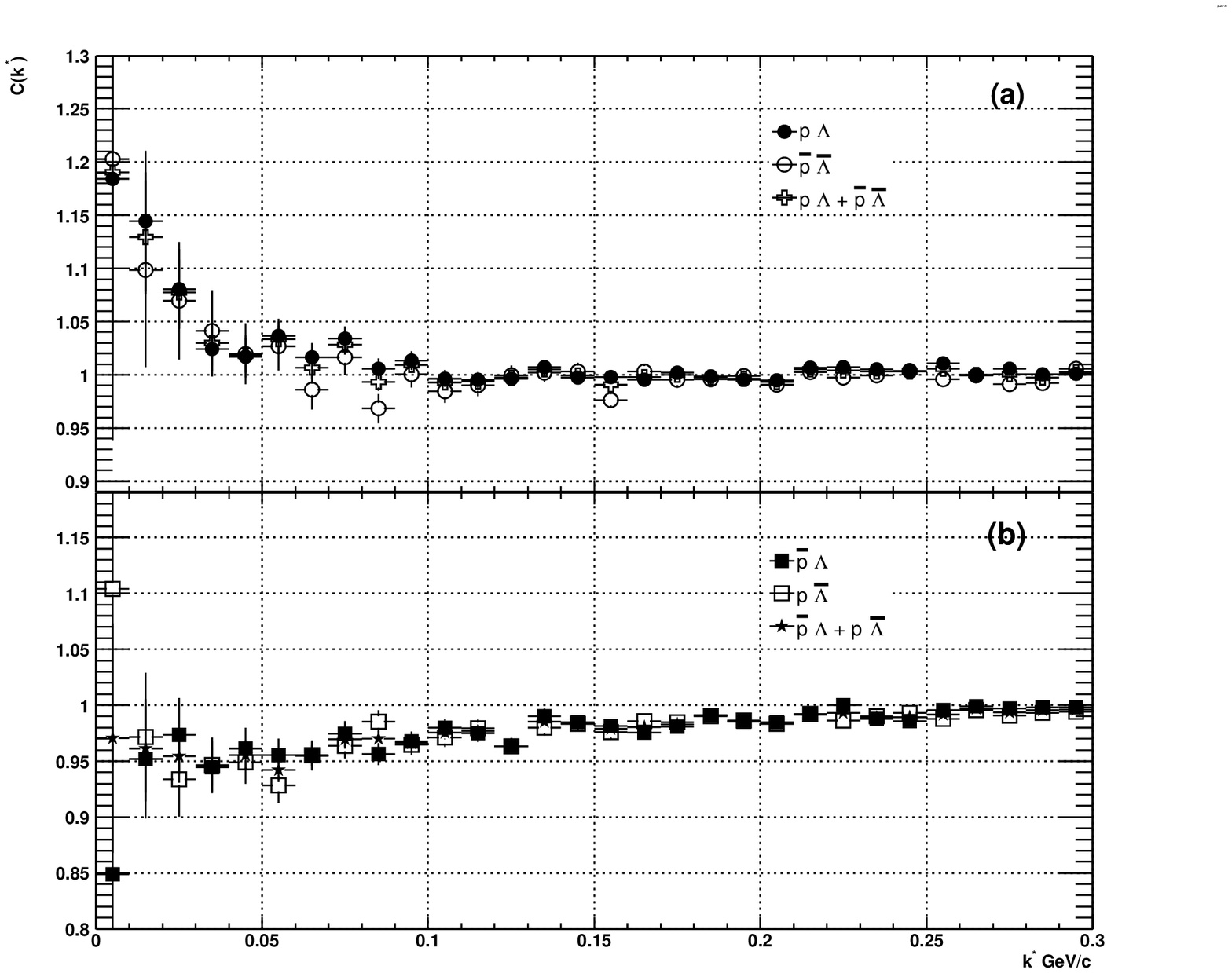}
\caption{The uncorrected measured $p-\Lambda$ correlation functions.
(a) $p-\Lambda$, $\overline{p}-\overline{\Lambda}$ and their sum.
(b) $\overline{p}-\Lambda$, $p-\overline{\Lambda}$ and their sum.} 
\label{pla}
\end{figure}

\begin{figure}[hbt!]
\center
\insertplot{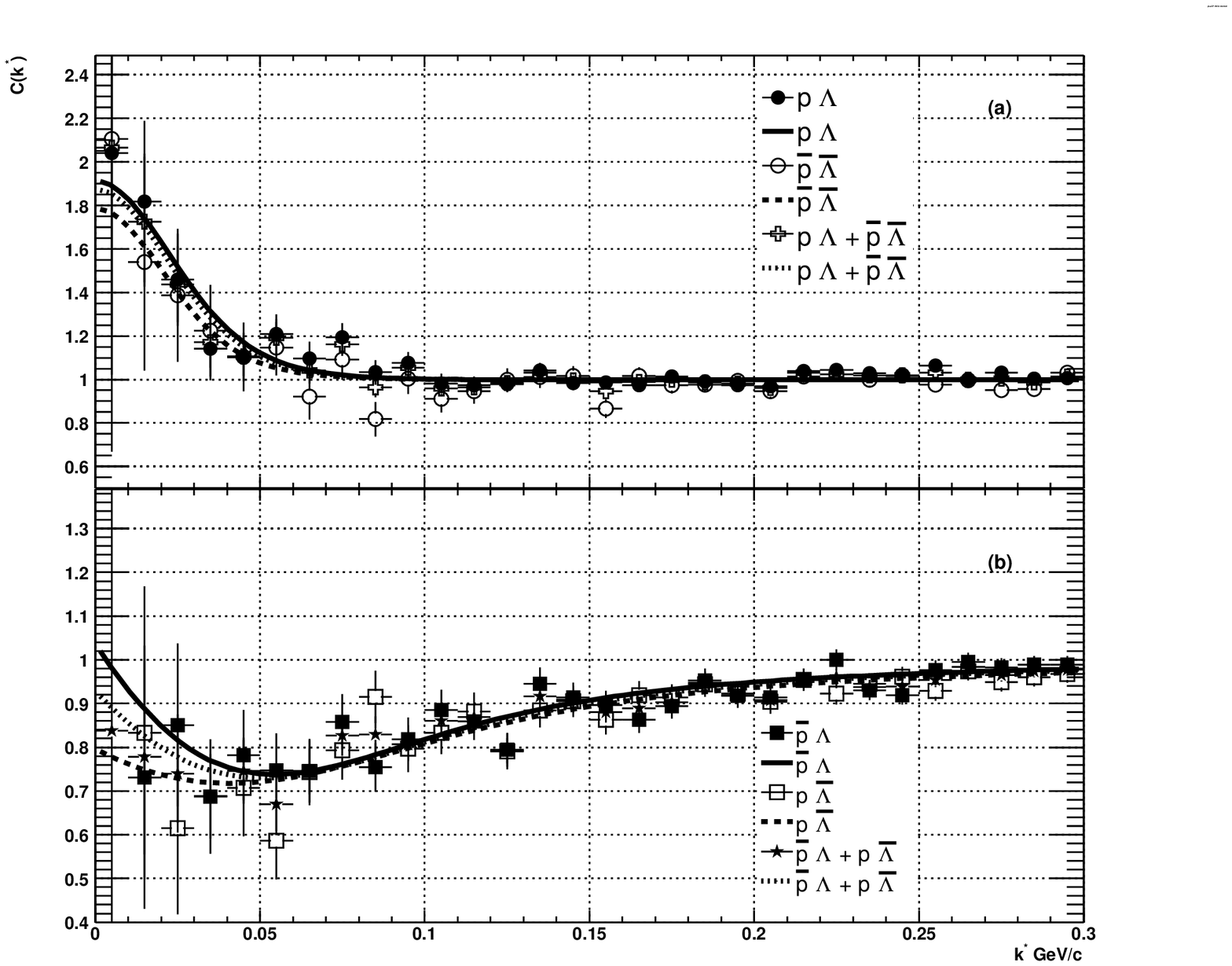}
\caption{The $p-\Lambda$ correlation functions corrected for purity and momentum resolution.
(a) $p-\Lambda$, $\overline{p}-\overline{\Lambda}$ and their sum.
(b) $\overline{p}-\Lambda$, $p-\overline{\Lambda}$ and their sum.
Curves correspond to fits done using the 
Lednick\'y \& Lyuboshitz analytical model.} 
\label{plaCorrFit}
\end{figure}

In addition, the effects of momentum resolution have been studied
using mixed pairs and by calculating the weight
with the Lednick\'y \& Lyuboshitz analytical model. It appears
that compared to statistical and systematic errors, 
the impact of the momentum resolution effect is negligeable.
Nevertheless, correlation functions have been corrected 
for the momentum resolution using the following formula:
\begin{eqnarray} 
C_{true}(k^*) = \frac{C_{measured}(k^*)*C_{Th-not-smeared}(k^*)}{C_{Th-smeared}(k^*)}
\end{eqnarray}
where $C_{true}(k^*)$ represents the corrected correlation function.

The ratio $C_{Th-not-smeared}(k^*)/C_{Th-smeared}(k^*)$ is the correction factor,
where  $C_{Th-not-smeared}(k^*)$ is estimated without taking into account 
the effect of momentum resolution and $C_{Th-smeared}(k^*)$ includes effecs
of the momentum resolution.


The $p-\Lambda$, $\overline{p}-\overline{\Lambda}$ correlation functions
and their sum are presented in 
Fig.\ref{pla} (a). Fig. \ref{pla} (b) represents the $\overline{p}-\Lambda$,
$p-\overline{\Lambda}$ correlation functions and their sum.
The $\overline{p}-\Lambda$ and
$p-\overline{\Lambda}$ correlation functions, measured for the first time, 
appears to be negative.

\begin{table}[hbt!]
\begin{center}
\begin{tabular}{|c|c|c|c|c|c|}
 \hline
  Systeme & Parameter & Value   \\ \hline
  &$f^{S}_{0}$  (fm) & 2.88  \\ \hline 
  &$d^{S}_{0}$  (fm) & 2.92  \\ \hline 
  &$f^{T}_{0}$  (fm) & 1.66  \\ \hline 
  &$d^{T}_{0}$  (fm) & 3.78  \\ \hline 
  $p-\Lambda$ & $r_{0}$  (fm) & $2.94 \pm 0.34 ^{+0.92}_{-1.45}$ \\ \hline
  $\overline{p}-\overline{\Lambda}$ & $r_{0}$  (fm) 
& $3.24 \pm 0.59 ^{+0.90}_{-1.14}$ \\ \hline
  $p-\Lambda$+$\overline{p}-\overline{\Lambda}$ & $r_{0}$  (fm) 
& $3.09 \pm 0.30 ^{+0.85}_{-1.41}$ \\ \hline
\end{tabular}
\caption{Parameters of $p-\Lambda$ and $\overline{p}-\overline{\Lambda}$
interaction used to determine the radius of the source of particles.}
\label{pLaParam}
\end{center}
\end{table}

 \begin{table}[hbt!]
\begin{center}
\begin{tabular}{|c|c|c|c|c|c|}
 \hline
 System & Parameter & Value (fm)   \\ \hline
 &$d_{0}$ (fm) & 0.0	\\ \hline
 $\overline{p}$ - $\Lambda$ &Im($f_{0}$)  & $1.88_\pm 1.78 ^{+2.74}_{-1.89}$ \\ \hline
 $\overline{p}$ - $\Lambda$ &Re($f_{0}$)  & $-2.82 \pm 1.28 ^{+2.16}_{-1.18}$\\ \hline
 $\overline{p}$ - $\Lambda$ &$r_{0}$  & $ 1.56 \pm 0.07 ^{+0.45}_{-0.97}$    \\ \hline
 $p$ - $\overline{\Lambda}$ &Im($f_{0}$) & $0.37_\pm 0.66 ^{+1.26}_{-0.38}$ \\ \hline
 $p$ - $\overline{\Lambda}$ &Re($f_{0}$)  & $-1.20 \pm 1.07 ^{+0.56}_{-0.54}$\\ \hline
 $p$ - $\overline{\Lambda}$ &$r_{0}$  & $ 1.41 \pm 0.10 ^{+0.45}_{-0.85}$    \\ \hline
 $\overline{p}$ - $\Lambda$ + $p$ - $\overline{\Lambda}$ &Im($f_{0}$)  & $1.01 \pm 0.92 ^{+2.43}_{-1.11}$ \\ \hline
 $\overline{p}$ - $\Lambda$ + $p$ - $\overline{\Lambda}$ &Re($f_{0}$)  & $-2.03 \pm 0.96 ^{+1.37}_{-0.12}$\\ \hline
 $\overline{p}$ - $\Lambda$ + $p$ - $\overline{\Lambda}$ &$r_{0}$  & $ 1.50 \pm 0.05 ^{+0.44}_{-0.92}$    \\ \hline
\end{tabular}
\caption{Parameters extracted from $\overline{p}-\Lambda$, $p$ - $\overline{\Lambda}$ and their sum.}
\label{pbarLaParam}
\end{center}
\end{table}
\section{Lednick\'y \& Lyuboshitz analytical model}

The Lednick\'y \& Lyuboshitz analytical model
uses the strong FSI for $p-n$\cite{3,4}
in the frame of the effective range approximation
in order to calculate correlation functions.
Since the only interaction between the two baryons
is the strong one, like $p$ and $n$, this model has been used
to study experimental correlations \cite{10}.

The distribution of the relative positions is assumed to be gaussian:
\begin{eqnarray}
\vec{r^{*}} \sim e^{- \vec{ r^{*} }^{2} /4r_{0}^{2}}
\end{eqnarray}
$r_{0}$ being considered as the radius of the source.
We consider that particles are not polarized.
For $\overline{p}-\Lambda$ and $p-\overline{\Lambda}$,
 the spin dependence of the scattering length is neglected
$f ^{S}  =  f^{T}  =  f$, with S and T for singlet and triple states.
The effective range ($d_{0}$) is set to zero.
These assumptions allow
to better constrain fitted parameters. 
An extra parameter, 
$Im(f_{0})$ (the imginary part of the scattering length) 
should be introduced 
to take into account 
the baryon - anti-baryon annihilation channel.
The fit parameters from \cite{2} have been used 
for $p-\Lambda$, $\overline{p}-\overline{\Lambda}$ correlation functions
and for their sum
to extract values of the radius $r_{0}$
(Table \ref{pLaParam}) after corrections for purity
and momentum resolution.
The extracted source parameters are close to 
values obtained in measurements
performed by NA49 (CERN) collaboration in Pb+Pb collisions at 158 AGeV  \cite{5}
and by the E895 (AGS) experiment in Au+Au
collisions at 4, 6, and 8 AGeV \cite{6}.
In particular, using the correlation function allows
to estimate the final state interaction parameters for 
the $\overline{p}$ - $\Lambda$ and the $p-\overline{\Lambda}$
systems (Table \ref{pbarLaParam}).

One can notice that the value of the imaginary part of the 
scattering length obtained for $\overline{p}-\Lambda$ pairs (Table \ref{pbarLaParam})
is in agreement with the scattering length ($0.8$ fm) previously obtained
for $p-p$ pairs spin averaged \cite{7}.

The source radius extracted from $p-\Lambda$ and $\overline{p}-\overline{\Lambda}$
correlation functions,
is smaller than the one extracted from $\overline{p}-\Lambda$
and  $p-\overline{\Lambda}$ correlation functions. 

\section{Conclusion}\label{conclusion}

The $p-\Lambda$, $\overline{p}-\overline{\Lambda}$, $\overline{p}-\Lambda$,
$p-\overline{\Lambda}$ correlation functions have been shown.
Corrections for momentum resolution and particle purity have been taken into account.
The pair purity has a stronger effect on the correlation function
than momentum resolution and is an important source of systematic errors.
These uncertainties have been included in the fit procedure
to the parameters and taken into account as systematic errors.
It has been shown that by studying $p-\Lambda$ and $\overline{p}-\overline{\Lambda}$
one can estimate the size of the source of particles at freeze-out.
Final state interaction parameters, such as the scattering length,
can be extracted from $\overline{p}-\Lambda$ and $p-\overline{\Lambda}$ 
correlation functions.

The negative shape of the $\overline{p}-\Lambda$ correlation function
is consistent with the one extracted for the 
symetric system, $p-\overline{\Lambda}$.
The large range in $k^{*}$ of the correlation function is still not understood, and 
results in a very small value of the extracted radius.

\vfill\eject
\end{document}